\documentclass[twocolumn,prl,showpacs]{revtex4}

\usepackage{graphicx}
\usepackage{dcolumn}
\usepackage{float}
\usepackage{amsmath}

\newcommand{\comment}[1]{}
\newcommand{\cm}{cm$^{-1}$ }
\newcommand{\cmi}{cm$^{-1}$}

\begin{document}

\title{\boldmath The a-axis optical conductivity of detwinned ortho-II
YBa$_{2}$Cu$_{3}$O$_{6.50}$\unboldmath}

\author{J. Hwang$^{1}$, J. Yang$^{1}$, T. Timusk$^{1}$, S.G. Sharapov$^{1}$,
J.P. Carbotte$^{1}$, D.A. Bonn$^{2}$,
Ruixing Liang$^{2}$, and W.N. Hardy$^{2}$}

\email{hwangjs@mcmaster.ca}

\affiliation{${1}$Department of Physics and Astronomy, McMaster
University, Hamilton, ON L8S 4M1, Canada\\
$^{2}$Department of Physics and Astronomy, University of British
Columbia, Vancouver, BC V6T 1Z1, Canada}

\date{\today}



%
%
\begin{abstract}
The a-axis optical properties of a detwinned single crystal of
YBa$_{2}$Cu$_{3}$O$_{6.50}$ in the ortho II phase (Ortho II Y123,
$T_c$= 59 K) were determined from reflectance data over a wide
frequency range (70 - 42 000 \cmi) for nine temperature values between
28 and 295 K. Above 200 K the spectra are dominated by a broad
background of scattering that extends to 1 eV. Below 200 K a shoulder
in the reflectance appears and signals the onset of scattering at 400
\cmi. In this temperature range we also observe a peak in the optical
conductivity at 177 \cmi. Below 59 K, the superconducting transition
temperature, the spectra change dramatically with the appearance of
the superconducting condensate. Its spectral weight is consistent, to
within experimental error, with the Ferrell-Glover-Tinkham (FGT) sum
rule. We also compare our data with magnetic neutron scattering on
samples from the same source that show a strong resonance at 31
meV. We find that the scattering rates can be modeled as the combined
effect of the neutron resonance and a bosonic background in the
presence of a density of states with a pseudogap. The model shows that
the decreasing  amplitude of the neutron resonance with
temperature is compensated for by an increasing of the bosonic background
yielding a net temperature independent scattering rate at high
frequencies. This is in agreement with the experiments.
\end{abstract}


\maketitle
%

The complete phase diagram of the high temperature superconducting
(HTSC) cuprates is still under intense debate. The normal state,
particularly in the underdoped region, is dominated by a variety of
not-well-understood cross-over phenomena that may either be
precursors to superconductivity or competing states.  These
include the pseudogap~\cite{timusk99}, the magnetic
resonance~\cite{rossat-mignod91,mook98,fong99}, the anomalous
Nernst effect~\cite{xu00,ong04}, stripe
order~\cite{tranquada95,dumm02,lucarelli03,benfatto03} and
possible superconducting fluctuations~\cite{emery95,corson99}. The
situation is further complicated by the presence of disorder and
the practical considerations that lead to a situation where a
given cuprate is not investigated with all the available
experimental techniques. Ideally, one would like to have a system
where disorder is minimized and several experimental techniques
can be used with the same crystals. As a step in that direction we present
detailed a-axis optical data on the highly ordered ortho-II phase
of YBa$_{2}$Cu$_{3}$O$_{6.50}$ (Ortho II Y123) and compare these
data with recent results from magnetic neutron scattering and
microwave spectroscopy on crystals from the same source.

An important motivation for a comparison between transport
properties and the magnetic neutron resonance comes from the
observation that the carrier life time, as measured by infrared
spectroscopy, is dominated by a bosonic
mode~\cite{thomas88,puchkov96d,carbotte99,schachinger03,hwang04}
whose frequency and intensity, as a function of temperature and
doping level, tracks the inelastic magnetic resonance at 41 meV
with in-plane momentum transfer of
$(\pi,\pi)$~\cite{carbotte99,hwang04}. The magnetic resonance has
also been invoked to explain other self-energy effects such as the
kink in the dispersion of angle-resolved photo emission spectra
(ARPES)~\cite{norman98,eschrig02,johnson01,chubukov04} and as a
hump-peak structure in tunneling spectra~\cite{zasadzinsky01}.
These effects have also been attributed to the electron-phonon
interaction~\cite{lanzara01}. While the bosonic excitation may not
be the fundamental engine of
superconductivity~\cite{kee02,hwang04}, nevertheless it is
important to map out the regions in the phase diagram where it can
be found and correlate the various experiments that yield evidence
of its presence.

The YBa$_{2}$Cu$_{3}$O$_{7-x}$ (Y123) material has been one of the
most thoroughly studied of all the HTSC systems but, like most
cuprates, it suffers from disorder associated with the charge
reservoir layer, in the case of Y123, disordered oxygen chains.
However, at an oxygen doping level of $x=0.50$, an ordered ortho-II
phase occurs with alternating full and empty CuO chains, doubling the
unit cell along the a-direction, and yielding a well ordered
stoichiometric compound~\cite{liang00,yamani03}. A further advantage
of this system is the availability of very large crystals, suitable
for neutron scattering. A disadvantage of the Y123 system is that the
crystals are not easily vacuum-cleaved and as a result, surface
sensitive probes such as angle resolved photoemission or scanning
tunneling microscopy have been used less with Y123. In contrast
Bi$_{2}$Sr$_2$CaCu$_{2}$O$_{8}$ (Bi-2212) cleaves easily and unlike
Y123, can also be overdoped. Optical spectroscopy has the advantage of
working on both systems equally well and offers a bridge between the
ARPES surface sensitive experiments and the large-volume probe of
neutron scattering.

The paper is organized as follows. We start with a brief
introduction to the experimental method followed by a presentation
of the raw reflectance data and an analysis where we use the
extended Drude model to extract the various optical constants, in
particular the real and imaginary parts of the scattering rate. To
improve the accuracy of the data, we first remove known features
in the conductivity spectra such as the transverse optical phonons and
the delta-function response of the superconducting condensate. The
measured scattering rates are then compared with a theoretical
model.  We next discuss the overall results by comparing them with
data from other experiments on Ortho II Y123, in particular recent
neutron data\cite{stock03} on samples from the same source.

\section{Experimental Method and Results.}

The detwinned ortho-II YBa$_{2}$Cu$_{3}$O$_{6.50}$ sample used in
this study was grown by a flux method using BaZrO$_{3}$
crucibles~\cite{liang98}, annealed under pure oxygen gas flow at
760 $^{o}$C, and detwinned at 300 $^{o}$C by applying uni-axial
stress of 100 atm along the a-direction~\cite{liang00}. The
dimension of the sample is $1.5\times1\times0.2\times$ mm$^3$. The
nearly normal-incident reflectance of the sample was measured over
a wide frequency range (70 - 42000 \cmi) at nine temperatures
between 28 K and 295 K with a Bruker IFS 66v/S Fourier transform
spectrometer with linearly polarized light. A polished stainless
steel mirror was used as an intermediate reference to correct for
instrumental drifts with time and temperature. An {\it in situ}
evaporated gold film on the sample was the absolute reflectance
reference~\cite{homes93b}. The reflectance of the gold films was
in turn calibrated with a polished stainless steel sample where we
relied on the Drude theory and the dc resistivity as the ultimate
reference. An advantage of this technique is that it corrects for
geometrical effects of an irregular surface. The {\it in situ}
gold evaporation technique gives an absolute accuracy of the
reflectance of better than $\pm 0.5$~\%. For the reflectance data
above 14 000 \cm we used aluminum instead of gold as the coating
material.
%
%
\begin{figure}[t!]
  \vspace*{-2.0cm}%
  \centerline{\includegraphics[width=3.5in]{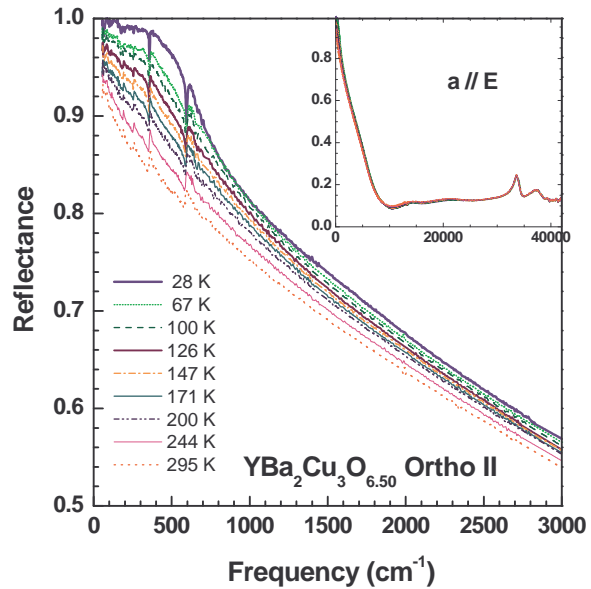}}%
  \vspace*{-1.9cm}%
\caption{The a-axis reflectance of Ortho II Y123 at nine
temperatures. There is a broad band of absorption giving rise to a
region of negative curvature of the reflectance between 350 and
600 \cmi. At high temperature the curvature is positive at all
frequencies. The inset shows the high frequency reflectance.}
\label{reflectance}
\end{figure}
%
%
\begin{figure}[t!]
  \vspace*{-2.0cm}%
  \centerline{\includegraphics[width=3.5in]{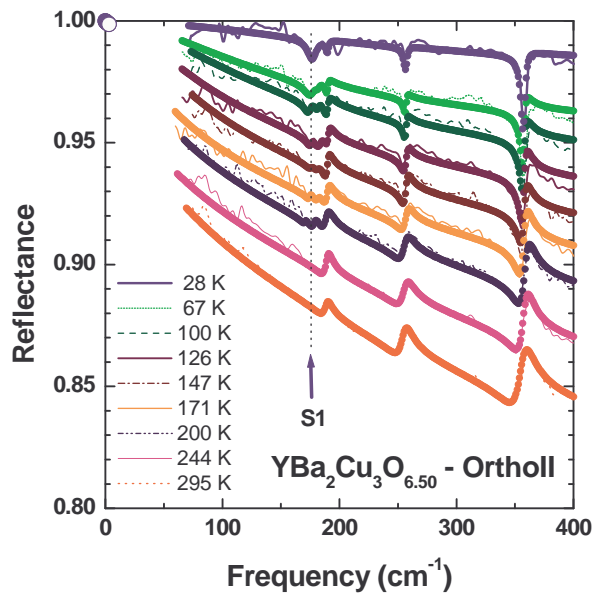}}%
  \vspace*{-1.9cm}%
\caption{The low frequency reflectance is shown as thin lines and an
oscillator fit is shown as heavy lines. The data are fitted with a
Drude band and a series of oscillators, one of which is a strong
electronic mode marked S1. In the superconducting state, an
oscillator at zero frequency is added to represent the
superconducting condensate. The oscillator parameters are shown in
Table I. The open circles near zero frequency are reflectance data
at 28 K from a microwave measurement~\cite{Pereg-Barnea04}.}
 \label{reffit}
\end{figure}
\begin{table}[htb]
  \begin{center}
  \begin{tabular}{|c|ccc|ccc|ccc|} \hline
&&28 K&&&100 K&&&295 K&\\\hline

&${\omega}_{pj}$&${\omega}_{j}$
&${\gamma}_{j}$&${\omega}_{pj}$&${\omega}_{j}$&${\gamma}_{j}$
&${\omega}_{pj}$&${\omega}_{j}$&${\gamma}_{j}$\\\hline\hline

Drude&6000&0&310&11 200&0&490&11 200&0&1500\\\hline

Extra&8500&0&1&5400&1&1&&&\\\hline

&430&190&3&500&190&4&225&189&7\\

Pho-&470&257&3.5&400&257&3&350&255&10\\

nons&720&358&4&690&358&3.5&520&356&13\\

&549&599&13&555&598&15&430&587&17\\\hline

S1&950&178&9&470&175&6&&&\\

&&&&355&182&4.5&&&\\\hline

\multicolumn{10}{|c|} {${\epsilon}_{\infty}$ = 3.63}\\\hline

\end{tabular}
  \end{center}
  \caption{The fitting parameters of Ortho II Y123 reflectance at three
temperatures, $T=$ 28 K, 100 K, and 295 K. The unit of
$\omega_{pj}$, $\omega_{j}$, and $\gamma_{j}$ is \cmi. An extra band
for $T= 28 $ K is the super fluid band at zero frequency. The extra band
for $T= 100 $ K shows additional spectral weight at low
frequency at low temperature (see Fig.~\ref{difference}). The peak S1
splits at 100 K into two components. }
  \label{table}
\end{table}

Figure~\ref{reflectance} shows the raw reflectance data for nine
different temperatures in the frequency region from 75 \cm to 3000
\cmi. The inset extends the data up to 42 000 \cmi. A strong
temperature dependence is observed in the low frequency range but
becomes weaker at higher frequencies. Two specific features show
temperature dependent amplitudes: a broad shoulder at $\approx$
400 \cm and a less obvious depression in reflectance at 175 \cmi.
In addition, signatures of well-known transverse optical phonons
can be seen as sharp minima in the low temperature
spectra~\cite{crawford89,tajima91,homes00}. At low temperature the
overall curvature of the reflectance is negative at low
frequencies changing to a positive curvature at high frequency,
which gives rise to an inflection point at $\approx$ 700 \cmi.
The inset shows the high frequency reflectance with a prominent
plasma edge at 13 000 \cm and a relatively weak temperature
dependence above 25 000 \cmi.

Figure~\ref{reffit} shows the lowest frequency range on an
expanded scale. In addition to the transverse phonons, we observe a
new feature at 175 \cmi, that appears below 171 K, and is denoted S1. The
thin lines are the measured reflectance curves and the solid thick
lines are oscillator fits where the phonons have been modeled as
Drude-Lorentz oscillators. The electronic background is
represented as a broad Drude peak in the normal state and a
two-fluid model with an additional delta function at the origin in
the superconducting state. The parameters used in the fit are
shown in Table I. From the fit we get a superconducting condensate
density of 72 $\times$ 10$^6$ cm$^{-2}$ expressed as the square of
a plasma frequency in \cmi. This corresponds to a penetration
depth of 1900 \AA. In comparison, the recent Gd Zero-field
Electron Spin Resonance (ESR) measurements on samples from the
same source yield a penetration depth of 2000 \AA\ and are shown as
open circles at very low frequency at 28 K~\cite{Pereg-Barnea04}.
Our estimated overall error in the absolute value of the
reflectance of $\pm$ 0.5 \% translates to a $\pm$ 300 \AA\ error
in penetration depth. We conclude that our data are in agreement
with the ESR results, consistent with errors associated with the
measurement of absolute reflectance. If we assume an overall
monotonic variation of the fitted curves with temperature, we can
estimate, from the deviation from this monotonicity, a relative
error of 0.2 \% for the temperature dependence of the reflectance.
Finally, from the noise level of the spectra, estimated to be 0.05
\% at 300 \cm and rising to 0.3 \% at 100 \cmi, we can set an
upper limit of 250 \cm to the plasma frequency of any phonon
features we could have missed above 200 \cmi. This limit would be
higher at lower frequencies.

The optical conductivity and the other optical constants were
determined from the measured a-axis reflectance by Kramers-Kronig
analysis~\cite{wooten72}, for which extrapolations to $\omega
\rightarrow 0 \:\:\mbox{and} \:\:\infty$ must be supplied. For
$\omega \rightarrow 0$, the reflectance was extrapolated by
assuming a Hagen-Rubens frequency dependence in the normal state,
$(1-R)\propto \omega^{1/2}$, and below $T_{c}$ an $(1-R)\propto
\omega^{4}$ extrapolation was used. We used the oscillator fits
shown in Fig.~\ref{reffit} to investigate the sources of error
arising from the low frequency extrapolations. The reflectance was
extended to high-frequency (between 40 000 and 350 000 \cmi) using
data from Romberg {\it et al.}~\cite{romberg90}. Free-electron
behavior ($R\propto \omega^{-4}$) was assumed to hold at higher
frequencies.
%
%
\begin{figure}[t!]
  \vspace*{-2.0cm}%
  \centerline{\includegraphics[width=3.5in]{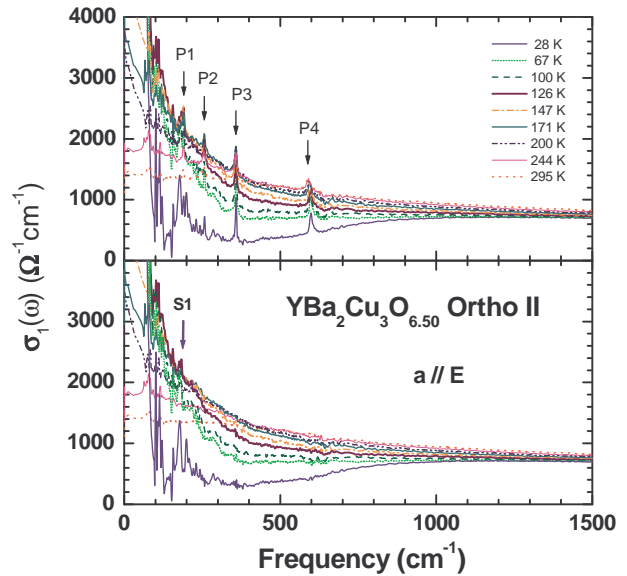}}%
  \vspace*{-1.9cm}%
\caption{The upper panel shows the optical conductivity where
several prominent phonon peaks have been identified as P1, P2 etc.
In the lower panel the phonon peaks (P1 to P4) have been removed.
A complex set of absorption bands can be seen at low temperature
dominated by a broad peak at 177 \cm designated as S1.}
\label{sigma}
\end{figure}

The top panel of figure~\ref{sigma} displays the optical
conductivity for the nine temperatures, eight in the normal state
and one in the superconducting state. The optical conductivity can
be written as
$\sigma(\omega)=-i\omega(\epsilon(\omega)-\epsilon_{H})/4\pi$
where $\epsilon_{H}$ is the dielectric constant at a high
frequency ($\sim$ 2 eV). At all temperatures we observe four sharp
phonon modes out of the six infrared active phonon
modes~\cite{crawford89,tajima91,homes00} expected for this
polarization of the incident radiation. The phonons have been
labelled P1 -- P4 in the figure. We fitted the optical
conductivity of these lines with a Drude-Lorentz model and found
the line intensities to be temperature independent and of a
magnitude expected for transverse optic (TO) phonons. Table I
shows the parameters of the model.

The bottom panel of figure~\ref{sigma} shows the optical
conductivity without the four phonon modes obtained through the
subtraction of the fitted Drude-Lorentz conductivities from the
measured conductivity. Several features stand out. First, there is
a prominent onset of conductivity that appears at low temperature
around 400 \cmi. This feature is common to all cuprate
superconductors and was identified early on as an onset of
scattering from a bosonic
mode\cite{thomas88,puchkov96d,marsiglio98}. Secondly, a strong
peak at $\approx$ 180 \cmi, designated S1, grows as the
temperature is lowered. At lower frequencies, below 150 \cmi,
there appears to be an additional strong absorption band, but any
structure here must be interpreted with caution since they are
derived from reflectance data close to unity which are subject to
large systematic errors. Finally, at higher frequencies, there is
a broad continuous background absorption that extends up to the
plasma frequency and has been attributed to the influence of
strong correlations~\cite{anderson97,millis90}.

\begin{figure}[t!]
  \vspace*{-2.0cm}%
  \centerline{\includegraphics[width=3.5in]{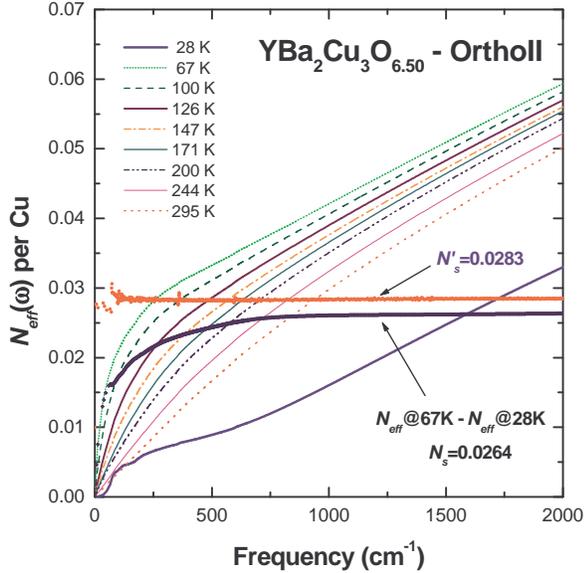}}%
  \vspace*{-1.9cm}%
\caption{The partial spectral weight up to 2000 \cm shown as
dashed and solid curves. The spectral weight increases
monotonically with temperature and frequency except for the 28 K
curve (heavy solid curve) in the superconducting state where there
is a loss of spectral weight to the condensate and the curve is
sharply lowered. The curve marked $N_s=N_{eff}(67 K)-N_{eff}(28
K)$ is an estimate of the spectral weight of the superconducting
condensate. The curve marked $N'_s$ is a plot of the condensate
density obtained independently from the imaginary part of
$\sigma(\omega)$.}%
\label{sumrule}
\end{figure}

We define the effective number of carriers per copper atom in
terms of the partial sum rule:
$N_{eff}(\omega)=\frac{2mV_{Cu}}{{\pi e^2}}\int_{0^+}^{\omega}
\sigma_1(\omega')d\omega'$ where $m$ is the free electron mass and
$V_{Cu}$ is the volume per copper atom, 57.7 \AA$^3$.
Figure~\ref{sumrule} shows $N_{eff}(\omega)$ calculated from the
optical conductivity. In the frequency region shown
$N_{eff}(\omega)$ increase uniformly with frequency and
temperature in the normal state.
%
%
%
%
\begin{figure}[t!]
  \vspace*{-2.0cm}%
  \centerline{\includegraphics[width=3.5in]{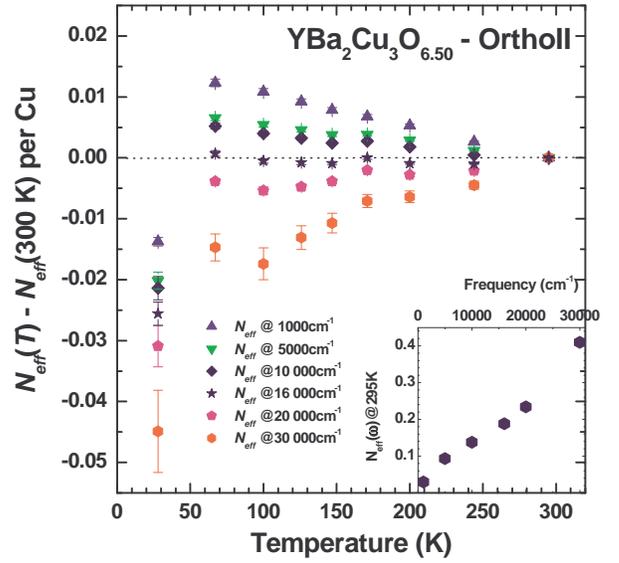}}%
  \vspace*{-1.9cm}%
\caption{The partial spectral weight integrated up to various
frequencies as a function of temperature. Below 16 000 \cm there
is an increase in spectral weight as the temperature is lowered
signaling a line narrowing on this frequency scale. Below $T_c$
there is strong loss of spectral weight to the superconducting
condensate. There is no evidence of any precursors to superconductivity
at 67 K. In the inset we show $N_{eff}(\omega)$ at 295 K.}%
\label{difference}
\end{figure}

In the superconducting state at 28 K, shown in
Figure~\ref{sumrule} as a heavy solid curve, there is a dramatic
loss of spectral weight due to the formation of the
superconducting condensate. The effective number of carriers per
copper atom in the condensate can be estimated from the partial
sum rule by subtracting the normal state curve at 67 K from the
superconducting curve at 28 K, also shown in the figure. The
resulting difference curve rises rapidly at low frequency and
saturates at a value of $N_s=0.0264$ above 1000 \cmi. To find a
better value of $N_s$ we need to estimate the normal state curve
at 28 K. We do this by extrapolating the temperature dependence in
the normal state to 28 K (see Fig.~\ref{difference}). With the
corrected value of $N_{eff}$ at 28 K we get a more accurate value
of $N_s=0.0276$. The condensate density can also be estimated from
the imaginary part of the conductivity provided the contribution
from the residual conductivity is not included {\it i.e.}
$N_s'=\frac{mV_{Cu}}{4
e^2}\omega[\sigma_2(\omega)-\sigma_{2r}(\omega)]=0.0283 $ where
$\sigma_{2r}$ is the residual quasiparticle conductivity in the
superconducting state. Figure~\ref{sumrule} shows both $N_s$ and
$N'_s$ and it is clear that the two curves (when corrected)
approach one another closely and the two methods differ by 2.5 \%
well within our 5 \% estimated experimental error. There have been
several reports of ab-plane spectral weight changes on entering
the superconducting state, in other high $T_c$
cuprates~\cite{tanner96,santander-syro02,molegraaf02} at the 1 to
2 \% level, but it should be noted that systematic errors in the
absolute value of reflectance have a large influence on the
magnitude of both $N_s$ and $N'_s$, although the error of the {\it
difference} is smaller. We estimate that our error in $N_s$ and
$N'_s$ is 30~\% but less than 5~\% in the difference $N'_s - N_s$.
Thus, at this point, we are unable to conclude from our data that
there is evidence for any added or missing low frequency spectral
weight when the superconducting state forms in the Ortho II Y123.

Figure~\ref{difference} shows the temperature dependence of the
partial spectral weight at various frequencies referred to the
spectral weight at 300 K.  We note that below 16 000 \cm the spectral
weight increases as the temperature is lowered while above this
frequency there is a decrease. At 16 000 \cm the partial spectral
weight is temperature independent. Since this frequency is close to
our estimate of the limit of the free carrier conductivity, we conclude
that there is a transfer of spectral weight to low frequencies as the
temperature is lowered within the free carrier band on two frequency
scales: a lower scale of the order of 500 \cm shown by the spreading
of the curves in Figure~\ref{sumrule}, and a higher one of the order
of 5000 \cm that is responsible for the temperature dependence below
10 000 \cm in Figure~\ref{difference}. We also note here that the
temperature dependence of the partial spectral weight shows no sign of
any precursors to superconductivity at 67 K which is 8 K above the
bulk superconducting transition temperature.

\comment{scattering rate}

The extended Drude model offers a detailed view of the charge
carrier scattering spectrum and its contribution to the effective
mass~\cite{allen71}. In this picture the scattering rate in the
Drude expression is allowed to have a frequency dependence:

\begin{eqnarray}
\sigma(\omega, T) &=& i \frac{\omega_p^2}{4 \pi} \frac{1}{\omega
+[\omega\lambda(\omega,T) + i/\tau(\omega, T)]} \nonumber\\  &=& i
\frac{\omega_p^2}{4 \pi} \frac{1}{\omega-2\Sigma^{op}(\omega, T)}
\end{eqnarray}
where $\omega_p$ is the plasma frequency, $1/\tau(\omega,T)$ is the
scattering rate and $\lambda(\omega)+1=m^*(\omega)/m$, $m^*(\omega)$
is an effective mass and $m$ the bare mass. We also introduce the
optical self energy $\Sigma^{op} \equiv \Sigma^{op}_1 + i
\Sigma^{op}_2$, where $-2\Sigma^{op}_1=\omega\lambda(\omega,T)$ and
$-2\Sigma^{op}_2=1/\tau$. The optical self energy is, apart from a
$cos(\theta)-1$ factor, where $\theta$ is a scattering angle, an
average over the Fermi surface of the quasiparticle
self-energy\cite{kaminski00,schachinger03,millis03,hwang04} as
measured by ARPES. Contrary to expectations, in optimally doped
Bi-2212 where both optical and ARPES data exist, the self-energies
derived from the two spectroscopies\cite{carbotte05a} are surprisingly
similar to one another~\cite{norman98,carbotte99,hwang04}.
%
%
\begin{figure}[t!]
  \vspace*{-2.2cm}%
  \centerline{\includegraphics[width=3.5in]{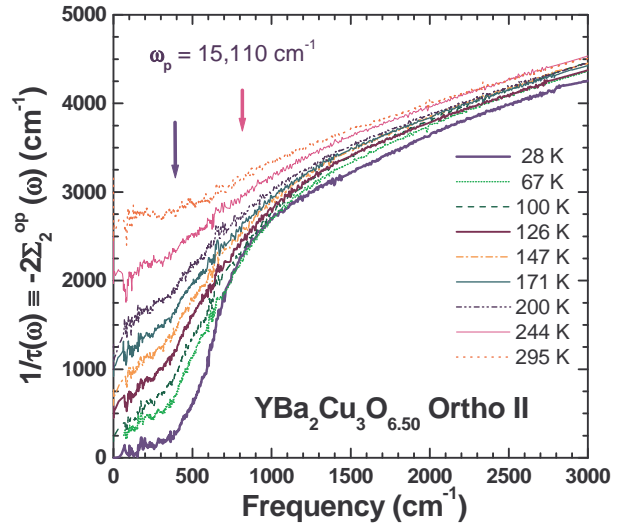}}%
  \vspace*{-2.0cm}%
\caption{The optical scattering rate obtained from the extended
Drude model. Two onsets, denoted by arrows dominate the scattering.}%
\label{tau}
\end{figure}
%
%
\begin{figure}[t!]
  \vspace*{-1.8cm}%
  \centerline{\includegraphics[width=3.5in]{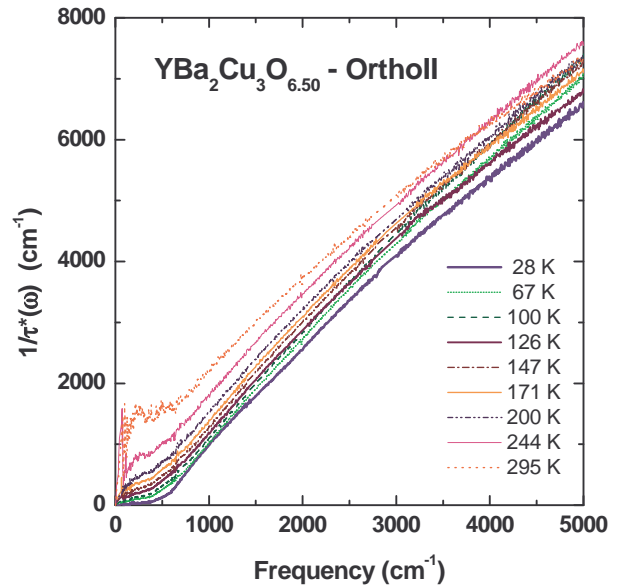}}%
  \vspace*{-1.5cm}%
\caption{The optical effective scattering rate. Given by the ratio
of the real and imaginary parts of the conductivity times the
frequency, this quantity has a linear frequency dependence that
extends to very high frequencies and a temperature dependence that
does not vary much with frequency.}%
\label{taus}
\end{figure}

\begin{figure}[t!]
  \vspace*{-2.0cm}%
  \centerline{\includegraphics[width=3.5in]{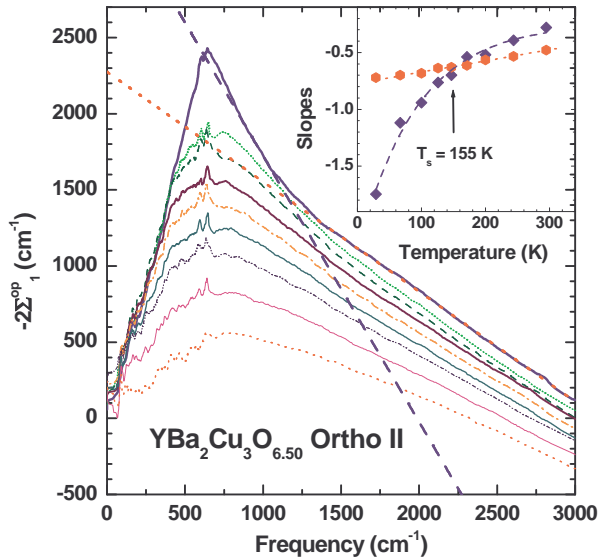}}%
  \vspace*{-1.9cm}%
\caption{The optical self-energy. The imaginary part of the
scattering rate is plotted. A broad peak is seen at 295 K (the
lowest curve). As the temperature is lowered this peak splits into
two components with a sharp peak appearing below 175 K as shown by
the curvature change at this temperature. The inset shows a
curvature analysis of the self-energy where the slope of the
self-energy in the two regions, shown in the main panel as two
dashed lines, have been plotted, circles, high frequency and
diamonds, low frequency.}
\label{selfenergy}
\end{figure}

Figure~\ref{tau} shows the scattering rate calculated from the
extended Drude model where $1/\tau(\omega,T)\equiv
-2\Sigma^{op}_2(\omega, T)={\omega_p^2 \over
 4\pi} {\rm Re} \big( {1
  \over \sigma(\omega,T) } \big)$ for Ortho II Y123. We also evaluate
the real part of the optical self-energy, which is given by:
 $-2\Sigma^{op}_1(\omega, T) = \omega \lambda(\omega,T) =
-{\omega_p^2 \over 4 \pi} {\rm Im} \big( {1 \over \sigma(\omega,T)
}\big) - \omega$ and is shown in Figure~\ref{selfenergy}. For the
calculation of the optical self-energy and the scattering rate we
need a value for the plasma frequency which includes spectral
weight up to the interband transitions. Using a procedure adopted
in a previous study~\cite{hwang04a} we find a plasma frequency
$\omega_{p}$=15 110 \cmi.

In Fig.~\ref{tau} we see an overall increase in scattering both
with temperature and frequency. The frequency dependence is
monotonic with two thresholds, shown with arrows, where the
scattering rate undergoes a step-like increase, a prominent low
frequency one at 400 \cm and a weaker, higher frequency one at 850
\cmi. The variation of the scattering rate with temperature is
roughly linear  at low frequency, but becomes much weaker at high
frequency above the high frequency threshold. The scattering rate
is larger than the frequency at all temperatures except at 28 K in
the superconducting state, {\it i.e.} $\hbar/\tau > \hbar\omega$.

Fig.~\ref{taus} shows the {\it effective} optical scattering rate
defined as:
\begin{equation}
\frac{1}{\tau^*(\omega, T)}= \omega \frac{\sigma_1(\omega,
T)}{\sigma_2(\omega, T)}
\end{equation}
One advantage of this quantity is that to calculate it the plasma
frequency is not needed. This quantity is also more linear with
frequency than the optical scattering rate~\cite{baraduc96}.
%
%

In Fig.~\ref{selfenergy} we show the real part of the optical
self-energy. It is related by Kramers-Kronig transformations to
the scattering rate shown in Fig.~\ref{tau}. The separation of the
two components of scattering is less obvious in this plot but in
analogy with the Bi-2212 system where we have investigated the
detailed doping dependent data on this quantity~\cite{hwang04}, we
find that the bosonic mode gives rise to a peak on top of a
broader background. With increased underdoping in Bi-2212 the peak
becomes triangular and harder to resolve from the background and
is very similar in appearance to the Ortho II Y123 data shown
here. We can attempt to resolve the peak by focusing on the break
in slope that occurs between 1000 \cm and 1400 \cm in the
temperature range from 28 K to 171 K. Above 200 K the break in
slope cannot be resolved. The inset to Fig.~\ref{selfenergy} shows
a plot of the slopes of two straight dashed lines that have been
fitted to the experimental data above and
below the break frequency. We find that a plot of the two slopes
shows that they cross at 155 K, {\it i.e.} the second derivative
of the optical self-energy goes through zero at this temperature. This
is shown in the inset to Fig.~\ref{selfenergy}

Recent calculations~\cite{chubukov04} of the self-energy of the
charge carriers interacting with collective spin excitations
suggest that the self-energy acquires an S-shaped frequency
dependence in the presence of a sharp mode, whereas in the absence
of such a mode the curve has a monotonically negative second
derivative. From these observations we conclude, as a first
approximation, that the bosonic mode is confined to temperatures
below 155 K.
%
%

\begin{figure}[t!]
  \vspace*{-2.0cm}
  \centerline{\includegraphics[width=3.5in]{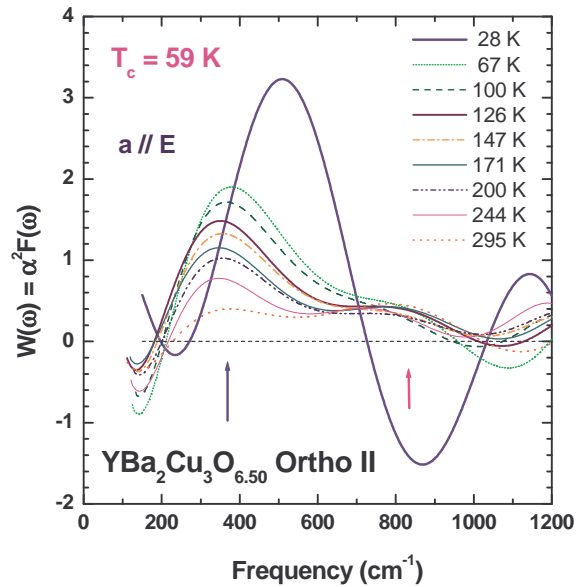}}
  \vspace*{-1.9cm}
 \caption{The Eliashberg function, $\alpha^2 F(\omega) \approx
W(\omega)$, which is calculated by using the Eq.~\ref{e08}. We
observe two peaks: one is temperature dependent at lower frequency
and the other is temperature independent at higher frequency. In
the superconducting state there is a large shift of the low energy
peak and a dramatic overshoot of the scattering rate giving rise
to a negative contribution to the Eliashberg function.}
  \label{resonance}
\end{figure}


For another estimate of the strength of the interaction of the
charge carriers with the sharp mode we used the procedure
introduced by Marsiglio {\it et
al.}~\cite{marsiglio98,carbotte99,abanov01} where the bosonic
spectral function is derived from the second derivative of the
optical scattering rate. This function can be written as
follows~\cite{allen71,marsiglio98,tu02}:

\begin{equation}
W(\omega)\equiv\frac{1}{2\pi}\frac{d^2}{d\omega^{2}}
\bigg[\frac{\omega}{\tau(\omega)}\bigg] \label{e08}
\end{equation}
and $W(\omega)\approx \alpha^2 F(\omega)$ at zero temperature in the
normal state, where $\alpha$ is a coupling constant, and $F(\omega)$
is a bosonic density of states. The results are shown in
Fig.~\ref{resonance}. The normal state shows two broad peaks; a
prominent one at $\approx$ 350 \cm and a much weaker one at $\approx$
800 \cmi. The higher frequency peak does not change with temperature
while the lower frequency one grows monotonically as temperature
decreases. Another interpretation is in terms of a temperature
dependent peak and a temperature independent background, four times
lower in amplitude. In the superconducting state there is a dramatic
shift of the lower peak to higher frequency and the development of a
region of negative spectral function $W(\omega)$ between 700 and 1000
\cmi. This behavior has been predicted by Abanov {\it et
al.}~\cite{abanov01}.

%
%
\begin{figure}[t!]
  \vspace*{-2.0 cm}
  \centerline{\includegraphics[width=3.5in]{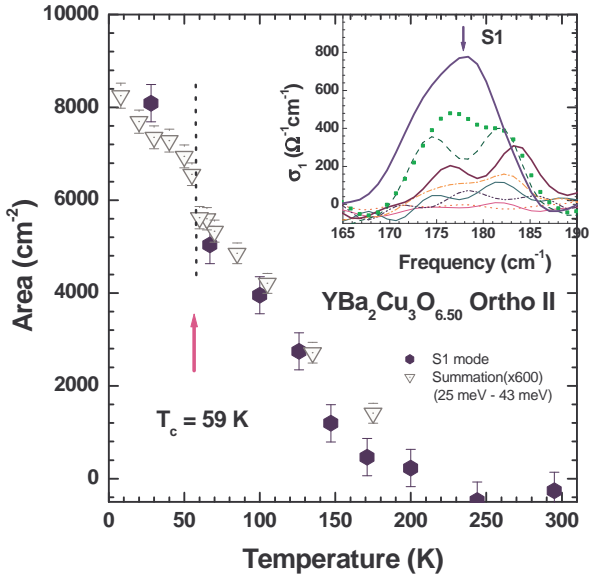}}%
  \vspace*{-1.9cm}
\caption{Area under the S1 mode as a function of temperature. We
also show the area under the neutron resonance peak~\cite{stock03}
at $Q=(1/2,3/2,2.2)$ and 33.1 meV, solid triangles from 25 to 43
meV (open triangles) as a function of temperature.}%
\label{S1mode}
\end{figure}

Fig.~\ref{S1mode}~ shows the area under the peak S1 as a function
of temperature obtained from a fit of a Lorentzian function to the
conductivity peak. We see that the intensity of this feature
decreases linearly with temperature going to zero at $\approx$ 200
K. On the same graph we have plotted the area under the neutron
resonance and it is clear that the two phenomena have parallel
temperature dependencies. We note that the spectral weight of the
S1 peak, 8000 cm$^{-2}$ represents only 0.006 \% of the total free
carrier spectral weight, 1.25x10$^8$ cm$^{-2}$.
%
%

\begin{figure}[t!]
  \vspace*{-2.0 cm}%
  \centerline{\includegraphics[width=3.5in]{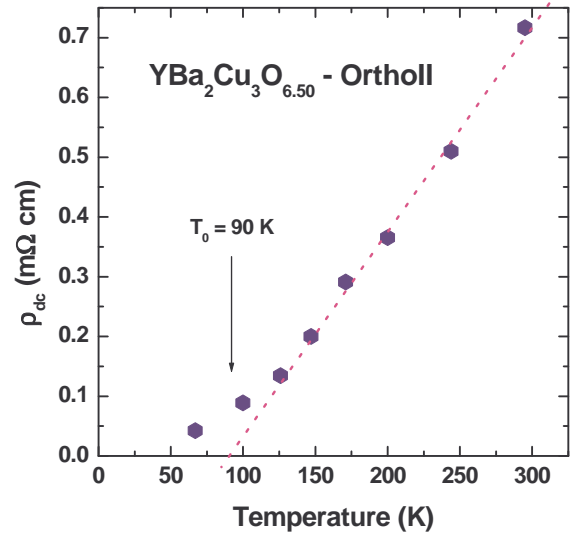}}%
  \vspace*{-1.9cm}%
\caption{DC resistivity extracted from the optical conductivity by
extrapolating the optical conductivity to zero frequency. The
temperature dependence shows the familiar linear variation with an
intercept on the temperature axis of $\approx$ 90 K.}%
\label{DCresist}
\end{figure}

We note that to obtain a finite frequency absorption peak such as
the peak S1, we need to break translational symmetry of the charge
carriers. We can rule out  phonon absorption which would be
inconsistent with the large spectral weight of the peak S1 and the
fact the spectral weight is temperature dependent as shown in
Fig~\ref{S1mode}. There are a number of possible candidate
mechanisms but we would like to focus on the static charge-density
patterns seen by scanning probe microscopy in
Bi-2212~\cite{vershinin04,mcelroy04}. Models for optical
absorption by strip-like patterns have been described by Benfatto
and Morais Smith~\cite{benfatto03}. The model describes the
optical absorption by charges confined by a static pinning
potential with two parameters, the strength of the potential and
the modulation amplitude.

Finally, we show the calculated dc resistivity extracted from the
extrapolated zero frequency limit of the optical conductivity.
Figure~\ref{DCresist} shows the resulting dc resistivity for the
eight normal state temperatures. The resistivity shows the
familiar linear temperature dependence with an intercept on the
temperature axis of 90 K, typical of underdoped samples of high
$T_c$ cuprates. A more detailed comparison shows some
discrepancies with the best dc transport data~\cite{ando04}.
First, our absolute value of the resistivity at 300 K is 0.7
(m$\Omega$cm)$^{-1}$ while Ando {\it et al.} find, at the same
doping level, a resistivity of 1.0 (m$\Omega$cm)$^{-1}$. The second
difference is the value of the resistivity in the low temperature
plateau region just above $T_c$ where Ando {\it et al.} measure a
residual resistivity of about 200 ($\mu\Omega$cm)$^{-1}$ whereas
our extrapolation yields less than 50 ($\mu\Omega$cm)$^{-1}$. We
think this difference may originate in the more ordered chain
layer in our samples which were annealed whereas the sample of
Ando {\it et al.} were quenched and presumably more disordered.
The lower absolute value of the 300 K resistivity is harder to
understand since we don't expect inelastic scattering, a mechanism
responsible for the high temperature resistivity, to depend on
disorder. One often suspects contact geometry in dc transport but
as Ando {\it et al.} have shown, errors from those sources amount
to at most a few percent in their data. Where the two data sets
are in good agreement is the value of $\approx$ 90 K for the
intercept on the temperature axis.

\section{Discussion}

We now turn to a general interpretation of our data in terms of
models proposed for the electrodynamics of cuprates focusing our
attention on the low frequency spectral region below the free
carrier plasma edge $\approx$ 10 000 \cmi. It is clear from even
the most superficial view of the data that the optical properties
shown here are not those of a set of free carrier interacting
weakly with impurities through elastic scattering or with lattice
vibrations through inelastic scattering. Two approaches have been
used to account for these deviations, the one and the two
component models of conductivity. In the one component model any
deviation from the simple Drude formula are attributed to a single
set of charge carriers interacting with a variety of bosonic
excitations while in the two component model a low frequency Drude
component is separated by curve fitting and any remaining
absorption is attributed to a mid-infrared band.

There is ample evidence that many conducting oxides do have a
separate well-defined mid-infrared band with a peak frequency that
moves to lower frequencies with doping~\cite{tokura91}. This is
also true of the one-layer cuprates, in particular
La$_{2-x}$Sr$_x$CuO$_4$~\cite{tanner87}. In some materials, when
the peak reaches zero frequency, striking changes occur in the
transport properties, for example, Ba$_{1-x}$K$_x$BiO$_3$ becomes
superconducting at this doping level~\cite{puchkov96a}. Recent
evidence from thermal conductivity suggests that
La$_{2-x}$Sr$_x$CuO$_4$ becomes an insulator below a doping level
of 0.06 where superconductivity also sets in whereas Y123 remains
in the conducting pseudogap state well below the doping
level~\cite{sutherland05} of the superconductivity boundary. These
observations suggest that the one-component to two-component
transition occurs at different doping levels for these two
materials and that the Y123 system remains metallic to the lowest
doping levels. We conclude that, at least at the level of doping
of ortho-II Y123, we are justified in using the one-component
model.

The striking result of the one-component model is the strong
frequency dependent scattering rate $1/\tau(\omega,T)$ shown in
Fig~\ref{tau}. There is a nearly perfect linear variation of
scattering at high frequency with a strong threshold at 400 \cm at
low temperatures that gradually washes with increasing
temperature. We will interpret our spectra within the
one-component model in terms of a scattering rate that is averaged
over the Brillouin zone and is the sum of two frequency dependent
terms, a bosonic mode that dominates at low temperature but whose
spectral weight weakens as the temperature rises and a featureless
temperature independent background that extends to high
frequencies.

Several theoretical models of strongly correlated system have been
evoked to explain the strong linearly rising frequency dependent
scattering rate. Earlier models included the Marginal Fermi
Liquid\cite{littlewood91}, the Nested Fermi Liquid
models~\cite{prelovsek01}, the Nearly Antiferromagentic
Liquid~\cite{millis90,monthoux93}, and the Luttinger Liquid
model~\cite{anderson97}. All models predict a continuous linear
rise of scattering governed by an energy scale of the order of
$\approx 0.5$ eV. The large energy scale rules out a simple phonon
mechanism for the broad background scattering which would predict
a flattening of the scattering rate beyond the maximum phonon
frequency~\cite{schachinger03}. The strongly temperature dependent
threshold has generally been interpreted in terms of coupling to a
bosonic mode~\cite{thomas88,puchkov96d}, in particular the the 41
meV magnetic resonance~\cite{carbotte99,schachinger03,hwang04}.

\subsection{Bosonic Mode Analysis of the Optical Scattering Rate}

Here we will adopt the approach of Schachinger {\it et
al.}~\cite{schachinger03} and treat both the mode and the
continuum scattering on an equal basis assuming that {\it both}
originate from coupling of the charge carriers to bosonic
fluctuations.

To study the temperature dependence of the coupling of bosonic modes to
charge carriers we used the following expression to model the scattering
rate within the extended Drude formalism:

\begin{equation}
\begin{split} \frac{1}{\tau(\omega,
T)}=&\frac{\pi}{\omega}\!\!\int^{+\infty}_{0}\!\!\!\!\!\!
d\Omega\alpha^2F(\Omega)\!\!\int^{+\infty}_{-\infty}
\!\!\!\!\!\!dz[N(z-\Omega)+N(-z+\Omega)]\\
&[n_B(\Omega)+1-f(z-\Omega)] [f(z-\omega)-f(z+\omega)]
\label{etau}
\end{split}
\end{equation}

where $\alpha^2F(\Omega)$ is the bosonic spectral
function~\cite{carbotte90}, $N(z)$ is the normalized density of
state of the quasiparticles, $n_B(\Omega)=1/(e^{\beta\Omega}-1)$,
$f(z)=1/(e^{\beta z}+1)$ are the boson and fermion occupation
numbers, respectively, and $\beta=1/(k_BT)$.

Eq.~\ref{etau} represents a finite temperature generalization of
the $T=0$ expression~\cite{mitrovic85},

\begin{equation}
\frac{1}{\tau(\omega)}=\frac{2
\pi}{\omega}\int_{0}^{\omega}d\Omega\alpha^2F(\Omega)\int_{0}^{\omega-\Omega}dz
\frac{1}{2}[N(z)+N(-z)] \label{mitrovic}
\end{equation}

Eq.~\ref{etau} is derived using the method proposed by Shulga {\it
et al.}~\cite{shulga91} and its derivation will be given
elsewhere~\cite{carbotte05}. Both Eq.~\ref{etau} and
~\ref{mitrovic} are suitable for the case when the quasiparticle
density of states, $N(z)$, cannot be taken as constant in the
vicinity of the Fermi levels, {\it e.g.} in the pseudogap state.

\begin{figure}[t!]
  \vspace*{-2.0 cm}%
  \centerline{\includegraphics[width=3.2in]{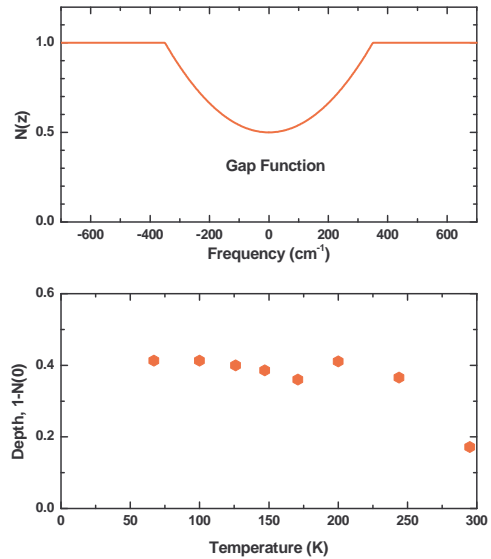}}%
  \vspace*{-1.9cm}%
\caption{The upper panel shows the quasiparticle density of state
with a gap, $N(z)$. The lower panel shows the depth of the
pseudogap for the least square fits. The other parameters of the
fit are shown in Table~\ref{table1}.}%
\label{gapf}
\end{figure}

The density of state $N(z)$ was modelled with a pseudogap with a
quadratic gap function:
\begin{equation}
N(z)=[N(0)+(1-N(0))\frac{z^2}{\Delta^2}]\theta(\Delta-|z|)+\theta(|z|-\Delta)
\end{equation}
where $1-N(0)$ is the depth of the gap, $\theta(z)$ is the
Heaviside step function, and $\Delta=350$ \cm is the frequency
width of the gap based on spectroscopic data for the pseudogap
from tunneling~\cite{kugler01}, and c-axis infrared
conductivity~\cite{homes93a}. An example for a depth of the gap of 0.50 is
shown in Fig.~\ref{gapf}.

\begin{figure}[t!]
  \vspace*{-2.0 cm}%
  \centerline{\includegraphics[width=3.2in]{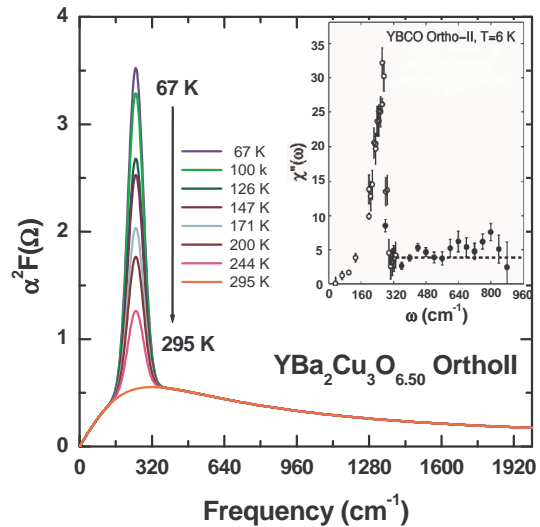}}%
  \vspace*{-1.9cm}%
\caption{The bosonic spectral function $\alpha^2 F(\Omega)$ obtained
from the least square fits to the experimental data.  In the inset we
show $\chi''(\omega)$ determined by neutron scattering from Fig. 14 in
Ref.~\cite{stock05}.}%
\label{a2F1}
\end{figure}

For the bosonic spectral function $\alpha^2F(\Omega)$ we used the
sum of two functions, a peak and a background.

\begin{eqnarray}
\alpha^2F(\Omega)&=&\mbox{PK}(\Omega)+\mbox{BG}(\Omega) \\
\mbox{PK}(\Omega)&=&\frac{A}{\sqrt{2\pi}(d/2.35)}e^{-(\Omega-\Omega_{PK})^2/[2(d/2.35)^2]}\\
\mbox{BG}(\Omega)&=&\frac{I_s \Omega}{\Omega_0^2+\Omega^2}
\end{eqnarray}
where PK$(\Omega)$ is a Gaussian peak and BG$(\Omega)$ is the
background which we have modelled on the spin fluctuation spectrum
of Millis, Monien, and Pines (MMP)~\cite{millis90}. $A$ is the
area under the Gaussian peak, $d$ is the full width at half
maximum (FWHM), and $\Omega_{PK}$ is its center frequency. We
fixed the parameters of the Gaussian peak to values shown in
Table~\ref{table1} based on the inelastic neutron scattering
results of Stock {\it et al.}~\cite{stock05}. $I_s$ is the
intensity of the MMP background and $\Omega_0$ is the frequency of
the background at maximum. The complete bosonic spectral function
$\alpha^2F(\Omega)$ used is shown in Fig.~\ref{a2F1}. It has two
adjustable parameters, the amplitude of the Gaussian peak $A$ and
depth of the pseudogap $1-N(0)$. All the other parameters  have
been determined from other experiments.

\begin{figure}[t!]
  \vspace*{-2.0 cm}%
  \centerline{\includegraphics[width=3.5in]{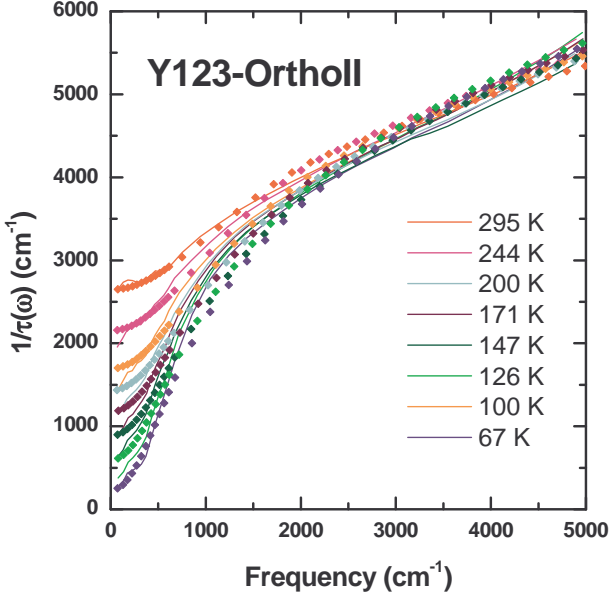}}%
  \vspace*{-1.90 cm}%
\caption{The optical scattering rates (solid lines) and our fits
(symbols) based on Eq.~\ref{etau} with the bosonic spectral
function shown in Fig.~\ref{a2F1} and pseudogap information of
Fig.~\ref{gapf} for the normal states. We note that the negative
temperature dependence of the mode contribution combined with the
almost equal and opposite contribution of the background results
in a nearly temperature independent scattering rate at high
frequencies, in complete agreement with the experimental results.}%
\label{Shulga}
\end{figure}

We used least squares to fit our scattering rate data shown in
Fig.~\ref{tau} to Eq.~\ref{etau}. The amplitude of the MMP
background and its center frequency were determined by fitting the
data at 295 K including only the MMP background in
$\alpha^2F(\Omega)$ with a very shallow gap (see
Table~\ref{table1}) in the Fermion density of states $N(z)$. We
fixed the background parameters for all other lower temperatures
to their 295 K values. For further fits at lower temperatures,
only two free parameters were used: the depth of the gap in the
density of states and the area under the resonance peak, $A$. The
calculated $1/\tau(\omega)$ spectra are compared with the measured
data in Fig~\ref{Shulga}. The dimensionless coupling constant or
mass enhancement factor $\lambda$ is defined as:
\begin{equation}
\lambda=2\int_{0}^{\infty} d\Omega
\frac{\alpha^2F({\Omega})}{\Omega}.
\end{equation}
The contributions to $\lambda_{PK}$ and $\lambda_{BG}$, from the
peak and the background as well as other parameters of the model
are shown in the Table~\ref{table1}.

%
%
%
%

One result of our fit is the temperature dependence of the depth
of the pseudogap shown in the lower panel of Fig.~\ref{gapf}. The
depth, we find, corresponds to a 41 \% depression of the density
of states at low temperature decreasing gradually to 17 \% at room
temperature. The depth of the pseudogap in the c-axis optical
conductivity is more pronounced~\cite{homes93a} but it is known
that c-axis transport is weighted more heavily in the antinodal
direction where the pseudogap is deeper whereas ab-plane transport
is more evenly distributed in momentum space.

Next, we compare our spectral function $\alpha^2F(\Omega)$ in
Fig.~\ref{a2F1} with the magnetic susceptibility $\chi''(\omega)$
determined by neutron scattering at 6 K ($<T_c$) by Stock {\it et
al.} (see inset in Fig.~\ref{a2F1}). The two sets of curves are
very similar although it should be pointed out that our data are
in the normal states and neutron data are in the superconducting
state and that the {\it width} of our mode has been set equal to
that of the neutron resonance mode.

Comparing the relative amplitudes of the peak and the background,
we find a substantial difference with the neutron data. For the
ratio of the background to the peak amplitude our
$\alpha^2F(\Omega)$ at 67 K gives $\sim$ 0.17 while the neutron
$\chi''(\omega)$ gives $\sim$ 0.27 in the normal state (we
estimated the ratio in the normal state from the temperature
dependent intensity of the neutron mode~\cite{stock03}). The
relative amplitude of the peak in our spectra is 63 \% stronger
than what is seen by neutron scattering. In the absence of a more
detailed calculation of the spectral function, this discrepancy is
not surprising since the coupling of the charge carriers to spin
fluctuations involves integrations over the Fermi surface and we
are reporting only on weighted averages. On the other hand our
data leaves open the possibility that the high frequency channel
of conductivity, that is active in producing our background, is
not well described by the MMP spin fluctuations. In addition to
the relative amplitudes we can also compare the areas under the
peak and background spectral functions. There is a sum rule for
the area under $\chi''(\omega)$ which is related to our optical
function $\alpha^2F(\Omega)$ by a coupling constant, $g^2$.
Neutron scattering finds that the area under the spin resonance is
about 3 \% of the total area. In our case at 67 K this fraction is
about 25 \%, considerably larger. It can be argued that the
coupling to the resonance $g^2$ could be larger than its
background value by a factor of $\sim \sqrt{8}$. This is
reasonable since the resonance is around ($\pi$,$\pi$) where the
susceptibility is also expected to peak.
%
%
\begin{figure}[t!]
  \vspace*{-1.0 cm}%
  \centerline{\includegraphics[width=3.5in]{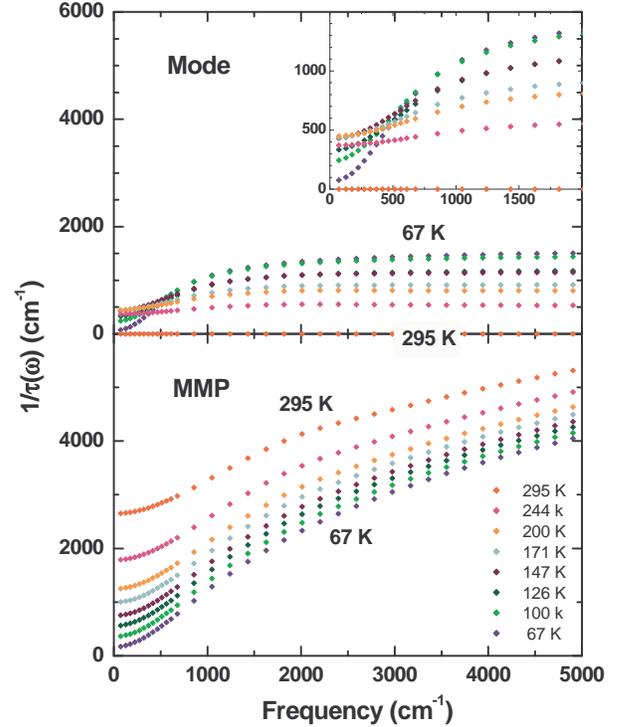}}%
  \vspace*{-0.50 cm}%
\caption{The mode contribution (upper panel) and the MMP back
ground contribution (lower panel) on the scattering rate based on
Eq.~\ref{etau} with the parameter shown in Table~\ref{table1} and
pseudogap information of Fig~\ref{gapf} for the normal states. We
note at high frequencies the opposite temperature dependencies of
the mode and the background contributions.}%
\label{modeMMP}
\end{figure}

%
%

\begin{table}[htb]
  \begin{center}
  \begin{tabular}{|c|cccc|ccc|c|} \hline
temp.&&peak&&&&MMP&&depth\\\hline

(K)&$A$&$\Omega_{PK}$&$d$&$\lambda_{PK}$&$I_s$&
$\Omega_{0}$&$\lambda_{BG}$&$1-N(0)$\\\hline\hline

67&255&248&80&2.20&354&320&3.30&0.413\\\hline

100&235&248&80&1.93&354&320&3.30&0.413\\\hline

147&170&248&80&1.40&354&320&3.30&0.386\\\hline

200&105&248&80&0.86&354&320&3.30&0.411\\\hline

244&62&248&80&0.51&354&320&3.30&0.366\\\hline

295&0&248&80&0.00&354&320&3.30&0.172\\\hline

\end{tabular}
  \end{center}
  \caption{The parameters of the bosonic mode analysis at six
representative temperatures, $T=$ 67 K, 100 K, 147 K, 200 K, 244
K, and 295 K. The peak is a Gaussian function (see Eq. 8) and MMP
is the background (see Eq. 9). The quantities $\lambda_{PK}$ and
$\lambda_{BG}$ are the coupling constants for the peak and the MMP
background, respectively. The depth, 1-$N(0)$, is the depth of the
gap at Fermi level in the density of state. All the frequencies
are measured in \cmi.}
  \label{table1}
\end{table}

%
%
\begin{figure}[t!]
  \vspace*{-2.0 cm}%
  \centerline{\includegraphics[width=3.5in]{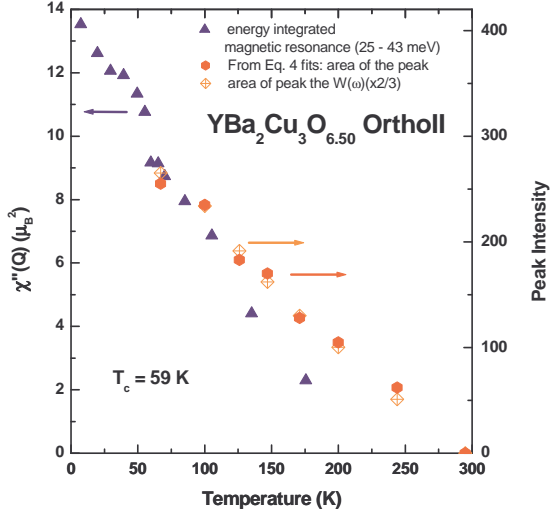}}%
  \vspace*{-1.9cm}%
\caption{Temperature dependence of the amplitude of the sharp
mode. The open diamonds with cross are from the second derivative
analysis of the scattering rate, the closed hexagons are from fit
a scattering rate to the model including a sharp mode and a
background. The upright triangles show the energy integrated
amplitude of the neutron mode Stock {\it et al.}~\cite{stock03}.}%
\label{comparison}
\end{figure}

In Fig.~\ref{comparison} we compare three quantities: the total
area under the magnetic susceptibility in the 25 to 43 meV energy
range from Stock {\it et al.}~\cite{stock03}, the area under the
peak at 350 \cm in the $W(\Omega)\approx \alpha^2F(\Omega)$ obtained
from the second derivative of $1/\tau(\omega)$, and the area under
the peak at 248 \cm in the $\alpha^2F(\Omega)$ from our fit. We
note that the two areas from the two different procedures for
deriving $\alpha^2F(\Omega)$ show almost identical temperature
dependencies. The area under the peak obtained from the magnetic
susceptibility shows a temperature trend similar to those of the
optical data.

\section{Discussion and conclusions.}

\subsection{The two channels of scattering of the charge carriers}

In analyzing the optical properties of Ortho II Y123 we have
adopted the one-component model of conductivity, attributing any
deviations from the simple Drude formula to inelastic processes.
We find that the high frequency processes are well described by
the MMP spectrum of spin fluctuations with a temperature
dependence that originates in the bosonic occupation numbers, but
a temperature {\it independent} spectral function, while at lower
frequencies a strong temperature dependence sets in and the
conductivity is well described by the extended Drude model with a
temperature {\it dependent} spectral function.

The two methods we used to estimate the shape and temperature
dependence of the bosonic spectrum gave very similar results. The
main contribution to the spectral weight came from a prominent
peak at 350 \cm which was strongly temperature dependent in
strength. From this temperature dependence we identified the mode
with the neutron resonance which has a frequency of 248 \cmi. The
100 \cm discrepancy we attributed to the effect of the pseudogap
in the density of states which, as calculations suggest, has the
effect of shifting upward the frequency of any bosonic mode. To
fit the data, the strength of the mode had to be temperature
dependent and the electronic density of states had to have a gap
whose depth was also temperature dependent. These observations are
in accord with what is known about the normal state of underdoped
Y123 from other experiments.

The properties of the sharp bosonic resonance in the infrared
scattering rate agree with the properties of the $(\pi,\pi)$
spin-flip neutron resonance mode. The frequency of the bosonic
resonance 248 \cm or 31 meV agrees with the frequency of the
neutron mode measured by Stock {\it et al.}~\cite{stock03}
(provided the pseudogap is included in the model). The temperature
dependence of the strength of the mode also agrees with the
neutron measurements on Ortho II Y123 samples from the same
source. The total spectral weight of the mode determined from
least squares fits drops monotonically with temperature and
reaches zero between 200 and 300 K, in agreement with the spectral
weight of the neutron mode.

Fig~\ref{modeMMP} shows separately the two contributions to the
scattering rate. We see that the combination of a sharp mode whose
strength decreases with temperature and a temperature independent
background solves a long standing problem in the optical
conductivity of the cuprates: there is a strong temperature
dependence at low frequencies but above 500 \cm the scattering is
nearly temperature independent. In the picture of mode and
background, both caused by coupling to bosons, this high frequency
behavior is the result of a competition between a rising
absorption due to occupancy of the background boson population and
a decrease in the strength of the bosonic mode with temperature.
This scenario predicts an overall linear temperature dependence at
all frequencies above the temperature where the sharp mode
disappears. There is some evidence for this kind of behavior in
the data for La$_{2-x}$Sr$_x$CuO$_4$ which does not have a
magnetic resonance but shows a linear temperature dependence
characteristic of bosons both in the dc
resistivity~\cite{gurvitch87} and optical scattering rate at high
frequency~\cite{startseva99}.

\subsection{Summary and conclusion}

In summary, we have shown that two optical phenomena in Ortho II Y123
have a close association with the magnetic resonance as measured for
samples from the same source. The first is the onset of scattering
associated with a bosonic resonance. This has been suggested by
several previous
investigators~\cite{carbotte99,johnson01,schachinger03, hwang04}, but
here, for the first time we base our conclusions on experiments that
use crystals with the same, well-defined doping level grown by the
same technique.

The second is the low-temperature conductivity peak that we have
designated S1. Its connection to the magnetic resonance is not obvious
since the magnetic resonance by itself is not optically active. One
possible mechanism is the growth of in-plane coherence that takes
place in this temperature region and has been suggested to be the
closely related to the transverse plasma resonance that is seen in
interplane conductivity~\cite{timusk03a}. The peak S1 may be such a resonance
associated with in-plane charge density modulations either through
pinning to defects~\cite{benfatto03} or through a transverse plasma
resonance associated with the inhomogeneous charge density that seems
to be present in underdoped cuprates in the normal
state~\cite{vershinin04,mcelroy04}.

%
%
\acknowledgments This work has been supported by the Canadian Natural
Science and Engineering Research Council and the Canadian Institute of
Advanced Research. We would like to thank Artem Abanov, Elena
Banscones, Dimitri Basov, Lara Benfatto, Andrey Chubukov, Chris Homes,
Peter Johnson, Alexandra Lanzara, Boza Mitrovic, Mike Norman, the late
Brian Statt, Setsuko Tajima, David Tanner, Tony Valla, and Zara
Yamani, for helpful discussions.

%
%


\begin{thebibliography}{99}

\bibitem{timusk99} For a review see: T. Timusk and B. Statt, Rep.
Prog. Phys. {\bf 62}, 61 (1999); an update to this review can be
found in T.~Timusk, Solid State Commun. {\bf 127}, 337 (2003).

\bibitem{rossat-mignod91} J. Rossat-Mignod, L.P. Regnault,
C. Vettier, P. Bourges, P. Burlet, J. Bossy, J.Y. Henry, and G.
Lapertot, Physica C {\bf 185-189}, 86 (1991). \comment{ spin gap
by neutron scattering L.P.~Regnault, C.~Vettier, P.~Bourges,
P.~Burlet, J.~Bossy, J.Y.~Henry, and G.~Lapertout, title: Neutron
scattering study of the YBa$_{2}$Cu$_{3}$O$_{6+x}$ system. First
observation of the 41 meV peak in the magnetic susceptibility}

\bibitem{mook98} H.A. Mook, Pengcheng Dai, S.M. Hayden, G. Aeppli,
T.G. Perring, and F. Do, Nature (London) {\bf 395}, 580 (1998).

\bibitem{fong99} H.F. Fong, P. Bourges, Y. Sidis, L.P. Regnault,
A. Ivanov, G.D. Gu, N. Koshizuka, and B. Keimer, Nature (London)
{\bf 398}, 588 (1999).

\bibitem{xu00} Z.A.~Xu, N.P.~Ong, Y. Wang, T.~Kakeshita, and
S.~Uchida, Nature (London) {\bf 406}, 486 (2000). \comment{Onset
of the vortex like Nerst signals above Tc in La2-xSrxCuO4 and
Bi2Sr2-yLayCuO6}

\bibitem{ong04} N.P.~Ong, Y.~Wang, S.~Ono, Y.~Ando, and
S.~Uchida, Ann. der Physik {\bf 13}, 9 (2004). \comment{Vorticity
and the Nernst effect in cuprate superconductors}

\bibitem{tranquada95} J.M.~Tranquada, B.J.~Sternlieb, J.D.~Axe,
Y.~Nakamura, and S.~Uchida, Nature (London) {\bf 375}, 561 (1995).
\comment{ first experimental paper on stripes}

\bibitem{dumm02} M.~Dumm, D.N.~Basov, S.~Komiya,~Y.~Abe, and
Y.~Ando, Phys. Rev. Lett. {\bf 88}, 147003 (2002).  \comment{LaSr,
stripes, ab plane,NdCe}

\bibitem{lucarelli03} A. Lucarelli, S. Lupi, M. Ortolani, P.
Calvani, P. Maselli, M. Capizzi, P. Giura, H. Eisaki, N. Kikugawa,
T. Fujita, M. Fujita, and K. Yamada, Phys. Rev. Lett. {\bf 90},
037002 (2003). \comment{ Phase Diagram of La2?xSrxCuO4 Probed in
the Infared: Imprints of Charge Stripe Excitations}

\bibitem{benfatto03} L.~Benfatto and C.M. Smith, Phys. Rev.
B {\bf 68}, 184513 (2003). \comment{ Signature of stripe pinning
in optical conductivity}

\bibitem{emery95} V.J. Emery and S.A. Kivelson, Nature (London) {\bf
374}, 434 (1995).  \comment{stripes model}

\bibitem{corson99} J.~Corson, R.~Malozzi, J,~Orenstein,
J.N.~Eckstein, and I.~Bozovic, Nature {\bf 398}, 221 (1999).
\comment{Normal state excess conductivity}.

\bibitem{thomas88} G.A. Thomas, J. Orenstein, D.H. Rapkine, M.
Capizzi,A.J. Millis, R.N. Bhatt, L.F. Schneemeyer, and J.V.
Waszczak, Phys. Rev. Lett. {\bf 61}, 1313 (1988).
\comment{"Ba2YCu3O7-d: Electrodynamics of Crystals with
HighReflectivity"}

\bibitem{puchkov96d} A.V.~Puchkov, D.N. Basov, and T. Timusk, J.
Physics: Condensed Matter, {\bf 8}, 10049 (1996). \comment{46
Pseudogap state in High $T_c$ Superconductors, an Infrared Study,}

\bibitem{carbotte99} J. P. Carbotte, E. Schachinger, and
D.N. Basov, Nature {\bf 401}, 354 (1999).

\bibitem{hwang04} J. Hwang, T. Timusk, and G.D. Gu. Nature (London)
{\bf 427}, 714 (2004).  \comment{High transition temperature
superconductivity in the absence of the mangetic resonance mode.}

\bibitem{schachinger03} E.~Schachinger, J.J.~Tu, and
J.P.~Carbotte,  Phys. Rev. B {\bf 67}, 214508 (2003).
 \comment{Angle-resolved photoemission spectroscopy and optical
 renormalizations:  Phonons or spin fluctuations}

\bibitem{norman98} M.R.~Norman and H. Ding, Phys. Rev. B {\bf 57},
R11089 (1998). \comment{Collective modes and the
superconducting-state spectral function of
Bi$_{2}$Sr$_{2}$CaCu$_{2}$O$_{8+x}$.[\pi,0 peak hump structure
related to a mode at 41 meV, suggested may be the neutron mode]}

\bibitem{eschrig02} M. Eschrig and M.R. Norman, Phys. Rev. B {\bf 67},
144503 (2003). \comment{Effect of magnetic resonance on the
electronic spectra of high Tc superconductors}

\bibitem{johnson01} P.D. Johnson, T. Valla, A.V. Fedorov,
Z. Yusof, B.O. Wells, Q. Li, A.R. Moodenbaugh, G.D. Gu, N.
Koshizuka, C. Kendziora, S. Jian, and D.G. Hinks, Phys. Rev. Lett.
{\bf 87}, 177007 (2001).

\bibitem{chubukov04} A.V. Chubukov and M.R. Norman, Phys. Rev. B {\bf 70},
174505 (2004).  \comment{Dispersion anomalies in cuprate
superconductors}

\bibitem{zasadzinsky01} J.F. Zasadzinski, L. Ozyuzer, N. Miyakawa,
K.E. Gray, D.G. Hinks, and C. Kendziora, Phys. Rev. Lett. {\bf
87}, 067005 (2001).

\bibitem{lanzara01} A. Lanzara, P.V. Bogdanov, X.J. Zhou, S.A. Kellar,
D.L. Feng, E.D. Lu, T. Yoshida, H. Eisaki, A. Fujimori, K. Kishio,
J.-I. Shimoyama, T. Noda, S. Uchida, Z. Hussain, Z.-X. Shen,
Nature (London) {\bf 412}, 510 (2001).

\bibitem{kee02} H.Y. Kee, S.A. Kivelson, and G. Aeppli, Phys. Rev. Lett.
{\bf 88}, 257002 (2002).

\bibitem{liang00} Ruixing Liang, D.A. Bonn, and Walter N. Hardy, Physica C
{\bf 336}, 57 (2000).

\bibitem{yamani03} Z. Yamani, W.A. MacFarlane, B.W. Statt, D. Bonn, R.
  Liang, and W.N. Hardy, cond-mat/0310255.
\comment{Cu NMR study of detwinned single crystals of Ortho-II
YBCO6.5}

\bibitem{stock03} C. Stock, W.J.L. Buyers, R. Liang, D. Peets, Z. Tun,
D. Bonn, W.N. Hardy, and R.J. Birgeneau, Phys. Rev. B {\bf 69},
014502 (2004).

\bibitem{liang98} Ruixing Liang, D.A. Bonn, W.N. Hardy, Physica C
{\bf 304}, 105 (1998).

\bibitem{homes93b} C.C. Homes, M.A. Reedyk, D.A. Crandles, and T. Timusk,
Appl. Opt. {\bf 32} 2976 (1993).

\bibitem{crawford89} M.K. Crawford, G. Burns, and F. Holtzberg,
Solid State Commun. {\bf 70}, 557 (1989).

\bibitem{tajima91} S. Tajima, T. Ido, S. Ishibashi, T. Itoh,
H. Eisaki, Y. Mizuo, T. Arima, H. Takagi, and S. Uchida, Phys. Rev. B
{\bf 43}, 10496 (1991).

\bibitem{homes00} C.C. Homes, A.W. McConnell, B.P. Clayman, D.A. Bonn,
Ruixing Liang, W.N. Hardy, M.Inoue, H. Negishi, P. Fournier, and
R.L. Greene, Phys. Rev. Lett. {\bf 84}, 5391 (2000).

\bibitem{Pereg-Barnea04} T. Pereg-Barnea, P.J. Turner, R. Harris, G.K. Mullins,
J.S. Bobowski, M. Raudsepp, Ruixing Liang, D.A. Bonn, and W.N.
Hardy, Phys. Rev. B {\bf 69}, 184513 (2004).

\bibitem{wooten72} Frederick Wooten, {\it Optical Porperties of Solids},
Academic, New York (1972).

\bibitem{romberg90} H. Romberg, N. N$\ddot{\mbox{u}}$cker, J. Fink, Th. Wolf,
X.X. Xi, B. Koch, H.P. Geserich, M. D$\ddot{\mbox{u}}$rrler, W.
Assmus, and B. Gegenheimer, Z. Phys B {\bf 78}, 367 (1990).

\bibitem{marsiglio98} F. Marsiglio, T. Startseva, and J.P. Carbotte,
Phys. Lett. A {\bf 245}, 172 (1998).

\bibitem{anderson97} P.W. Anderson, Phys. Rev. B {\bf 55} 11785
(1997). \comment{Infrared conductivity of cuprate metals, Detailed
fit using Luttinger-liquid theory}

\bibitem{millis90} A.J. Millis, H. Monien, and D. Pines, Phys. Rev.  B
{\bf 42}, 167 (1990).

\bibitem{tanner96} D.B.~Tanner, Y.-D.~Yoon, A.~Zibold, H.L.~Liu,
M.A.~Quijada, S.W.~Moore, J.M.~Graybeal, B.-H.O, J.T.~Markert,
R.J.~Kelley, M.~Onellion, and J.-H. Cho, in {\it Spectroscopic
Studies of Superconductors,} Ivan Bozovic, Dirk van der Marel,
Editors, Proc. SPIE 2696, 13, (1996).
\comment{doping and carrier density in cuprate superconductors}

\bibitem{santander-syro02} A.F. Santander-Syro, R.P.S.M. Lobo, N. Bontemps,
Z. Konstantinovic, Z.Z. Li, and H. Raffy, Phys. Rev. Lett. {\bf 88},
097005 (2002).

\bibitem{molegraaf02} H.J.A. Molegraaf, C. Presura, D. van der Marel,
P.H. Kes, and M. Li, Science {\bf 295}, 2239 (2002).

\bibitem{allen71} P.B. Allen, Phys. Rev. B {\bf 3}, 305
(1971).

\bibitem{kaminski00} A. Kaminski, J. Mesot, H. Fretwell, J. C. Campuzano,
M. R. Norman, M. Randeria, H. Ding, T. Sato, T. Takahashi,
T.~Mochiku, K. Kadowaki, and H. Hoechst Phys. Rev. Lett. {\bf 84},
1788 (2000). \comment{Quasiparticles in the superconducting state
of Bi$_{2}$Sr$_{2}$CaCu$_{2}$O$_{8+\delta}$}

\bibitem{millis03} A.J. Millis and H.D.~Drew, Phys. Rev. B {\bf
67}, 214517 (2003). \comment{title: Quasiparticles in high
temperature superconductors: consistency of angle resolved
photoemission and optical conductivity.}

\bibitem{carbotte05a} J.P. Carbotte, E. Schachinger, and J. Hwang
Phys. Rev. B {\bf 71}, 054506 (2005).

\bibitem{hwang04a} J. Hwang, T. Timusk, A.V. Puchkov, N. L. Wang,
G.D. Gu, C.C. Homes, J.J. Tu, and H. Eisaki, Phys. Rev. B {\bf
69}, 094520 (2004).

\bibitem{baraduc96} C. Baraduc, A. El Azrak, and N. Bontemps,
J. Supercond. {\bf 9}, 3 (1996).

\bibitem{abanov01} A.. Abanov, A.V. Chubukov, and J. Schmalian,
Phys. Rev. B {\bf 63}, 180510(R) (2001).


\bibitem{tu02} J.J. Tu, C.C. Homes, G.D. Gu, D.N. Basov,
and M. Strongin, Phys. Rev. B {\bf 66}, 144514 (2002).

\bibitem{vershinin04} M. Vershinin, S. Misra, S. Ono, Y. Abe, Y. Ando,
and A. Yazdani, Science {\bf 303}, 1995 (2004).

\bibitem{mcelroy04} K. McElroy, D.-H. Lee, J.E. Hoffman, K.M. Lang,
E.W. Hudson, H. Eisaki, S. Uchida,J. Lee, and J.C. Davis,
cond-mat/040405. \comment{Homogeneous nodal superconductivity
coexisting with inhomogeneous charge order in strongly underdoped
Bi2Sr2CaCu2O8+\delta}

\bibitem{ando04} Yoichi Ando, Seiki Komiya, Kouji Segawa, S. Ono, and Y.
Kurita, Phys. Rev. Lett. {\bf 93}, 267001 (2004).

\bibitem{tokura91} S. Uchida, T. Ido, H. Takagi, T. Arima, Y. Tokura,
and S. Tajima, Phys. Rev. B {\bf 43}, 7942 (1991).

\bibitem{tanner87} S.L. Herr, K. Kamaras, C.D. Porter, M.G. Doss, D.B.
Tanner, D.A. Bonn, J.E. Greedan, C.V. Stager, and T. Timusk, Phys.
Rev. B {\bf 36}, 733 (1987).

\bibitem{puchkov96a} A.V. Puchkov, T. Timusk, M.A. Karlow, S.L. Cooper,
P.D. Han, and D.A. Payne, Phys. Rev. B {\bf 54}, 6686 (1996).

\bibitem{sutherland05} M. Sutherland, S.Y. Li, D.G. Hawthorn, R.W. Hill,
F. Ronning, M.A. Tanatar, J. Paglione, H. Zhang, L. Taillefer, J.
DeBenedictis, Ruixing Liang, D.A. Bonn, and W.N. Hardy,
cond-mat/0501247.

\bibitem{littlewood91} P.B. Littlewood and C.M. Varma, J. Appl.
Phys. {\bf 69}, 4979 (1991).

\bibitem{prelovsek01} P. Prelovsek, T. Tohyama, and S. Maekawa, Phys. Rev. B
{\bf 64}, 052512 (2001).

\bibitem{monthoux93} P. Monthoux and D. Pines, Phys. Rev. B {\bf 47},
6069 (1993).

\bibitem{carbotte90} J.P.~Carbotte, Rev. Mod. Phys. {\bf 62}, 1027 (1990).

\bibitem{mitrovic85} B. Mitrovic and M.A. Fiorucci, Phys. Rev. B {\bf 31},
2694 (1985).

\bibitem{shulga91} S.V. Shulga, O.V. Dolgov, and E.G. Maksimov,
Physica C {\bf 178}, 266 (1991).

\bibitem{carbotte05} S.G. Sharapov and J.P. Carbotte, in preparation.

\bibitem{kugler01} M.~Kugler, \O. Fischer, Ch.~Renner, S.~Ono, and
Yoichi Ando, Phys. Rev. Lett. {\bf 86}, 4911 (2001).

\bibitem{homes93a} C.C. Homes, T. Timusk, R. Liang,  D.A. Bonn,
and W.N. Hardy, Phys. Rev. Lett. {\bf 71}, 1645 (1993).

\bibitem{stock05} C. Stock, W.J.L. Buyers, R.A. Cowley, P.S. Clegg, R. Coldea,
C.D. Frost, R. Liang, D. Peets, D. Bonn, W.N. Hardy, and R.J.
Birgeneau, Phys. Rev. B {\bf 71}, 024522 (2005).

\bibitem{gurvitch87} M. Gurvitch and A.T. Fiory,  Phys.
Rev. Lett. {\bf 59}, 1337 (1987).

\bibitem{startseva99} T. Startseva, T. Timusk, A. V. Puchkov, D.
N. Basov, H. A. Mook, M. Okuya, T. Kimura, and K. Kishio, Phys.
Rev. B {\bf 59}, 7184, (1999). \comment{Temperature evolution of
the pseudogap state in the infrared response of underdoped La2 -
xSrxCuO4 \LaSr}

\bibitem{timusk03a} T. Timusk and C.C. Homes, Solid State Comm. {\bf
  126,} 63 (2003).

\end{thebibliography}
\end{document}